\documentclass[10pt,a4paper,twoside]{article}
\usepackage{epsfig}
\usepackage{baltlat6}
\usepackage{array}
\usepackage{here}
\begin{document}
\ \
\newcommand{\ion}[2]{#1~{\small #2}}
\newcommand{\de}{\rm d}
\vspace{0.5mm}
\setcounter{page}{277}
\vspace{8mm}

\titlehead{Baltic Astronomy, vol.\,VV, ppp--ppp, 2011}

\titleb{OPTICAL EMISSION LINES AND THE X-RAY PROPERTIES OF TYPE 1 SEYFERT GALAXIES}

\begin{authorl}
\authorb{G. La Mura}{1}
\authorb{S. Ciroi}{1}
\authorb{V. Cracco}{1}
\authorb{D. Ili\'c}{2}
\authorb{L. \v{C}. Popovi\'c}{3}
\authorb{P. Rafanelli}{1}
\end{authorl}

\begin{addressl}
\addressb{1}{Department of Astronomy, University of Padua,\\
Vicolo dell'Osservatorio 3, Padova I-35122, Italy; giovanni.lamura@unipd.it, stefano.ciroi@unipd.it, valentina.cracco@unipd.it, piero.rafanelli@unipd.it}
%Institute of Theoretical Physics and Astronomy, Vilnius
%University,\\  Go\v{s}tauto 12, Vilnius LT-01108, Lithuania;
%straizys@itpa.lt}
\addressb{2}{Department of Astronomy, Faculty of Mathematics, University of Belgrade,\\ Studentski trg 16, Belgrade 11000, Republic of Serbia; dilic@matf.bg.ac.rs}
\addressb{3}{Astronomical Observatory of Belgrade,\\ Volgina 7, Belgrade 11060, Republic of Serbia; lpopovic@aob.bg.ac.rs}
\end{addressl}

\submitb{Received: 2011 June 10; accepted: 2011 MM DD}

\begin{summary} In this contribution we report on the study of the optical emission lines and X-ray spectra of a sample of Type 1 AGNs, collected at the {\it Sloan Digital Sky Survey} database and observed by the {\it XMM-Newton} satellite. Exploiting the different instruments carried onboard {\it XMM}, we identify the spectral components of the soft and hard energy bands (in the range from 0.3 keV up to 10 keV). The properties of the X-ray continuum and of the Fe K$\alpha$ line feature are investigated in relation to the optical broad emission line profiles and intensity ratios. The resulting picture of emission, absorption and reflection processes is interpreted by means of a BLR structural model that was developed on the basis of independent optical and radio observations.
\end{summary}


%% \resthead is the RUNNING TITLE at top of the pages
\resthead{Optical and X-ray properties of Type 1 Seyfert galaxies}
{G. La Mura et al.}

\sectionb{1}{INTRODUCTION}

The spectra of Type 1 Seyfert galaxies are characterised by prominent emission lines and by a strong continuum of ionizing radiation. In general, it is accepted that such spectral components descend from our direct view of the central engine of an Active Galactic Nucleus (AGN), where accretion of matter onto a Super Massive Black Hole results in the emission of large amounts of energy (typically $10^{41} {\rm erg\, s^{-1}} \leq L \leq 10^{46} {\rm erg\, s^{-1}}$). The mass of the SMBH can be estimated from the width of the broad emission lines, which are produced very close to the continuum source, in the so called Broad Line Region (BLR):
$$M_{BH} = f \frac{R_{BLR} v^2}{G}, \eqno(1)$$
where $v$ is an estimate of the plasma velocity (usually connected to the width of the emission lines), $R_{BLR}$ is the size of the region, $G$ is the gravitational constant, and $f$ is a geometrical factor, accounting for the influence of the source structure on the formation of the line profiles.

In this work we present a combined analysis of Type 1 AGN optical and X-ray spectra, devoted to the investigation of the intrinsic properties of the central engine and their influence on the BLR plasma.

\sectionb{2}{BLR PLASMA DIAGNOSTICS}

If we consider an emission line, originated by the transition from an upper level $u$ to a lower level $l$, we can express the associated intensity as:
$$I_{u l} = \frac{h c}{\lambda_{u l}} A_{u l} \int_{R_1}^{R_2} N_u(r)\cdot e^{-\tau(r)} \de r, \eqno(2)$$
where $A_{u l}$ is the probability of a spontaneous radiative decay, $N_u(r)$ the number density of the emitting ions, $\tau(r)$ is the optical depth of the layer, $h$, $c$, and $\lambda_{u l}$ represent the Planck constant, the speed of light, and the line wavelength, while $R_1$ and $R_2$ are the boundaries of the emitting region. In the case of an optically thin plasma, with the population of the high excitation energy levels following the Saha-Boltzmann distribution (plasma in a {\it Partial Local Thermodynamic Equilibrium}, or PLTE; see Popovi\'c 2003, 2006),  we can introduce a normalized line intensity, with respect to the atomic transition constants, and write Eq.~(2) in the form of:
$$I_n = \frac{\lambda_{u l} I_{u l}}{g_u A_{u l}} = h c N_0 \ell e^{-E_{u 0} / k_B T}. \eqno(3)$$
in which $g_u$ is the statistical weight of the upper level, $\ell$ is the spatial extension of the emitting region, $N_0$ is the number density of the ions in their unexcited configuration, while $E_{u 0}$, $k_B$, and $T$ are the excitation energy of the level, the Boltzmann constant, and the plasma electron temperature, respectively.

\begin{figure}[t]
\begin{center}
\includegraphics[width=6cm]{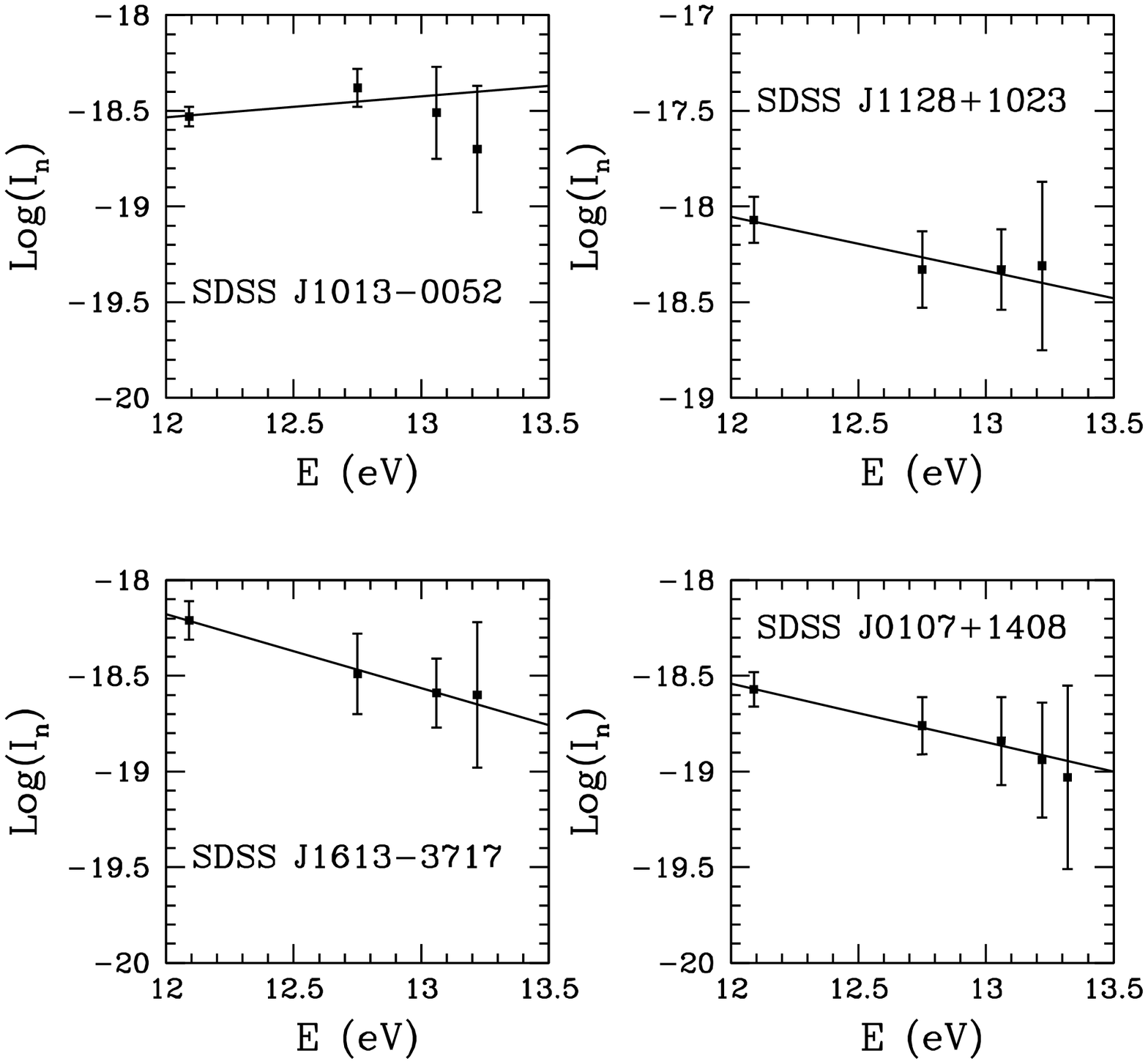}
\includegraphics[width=6cm]{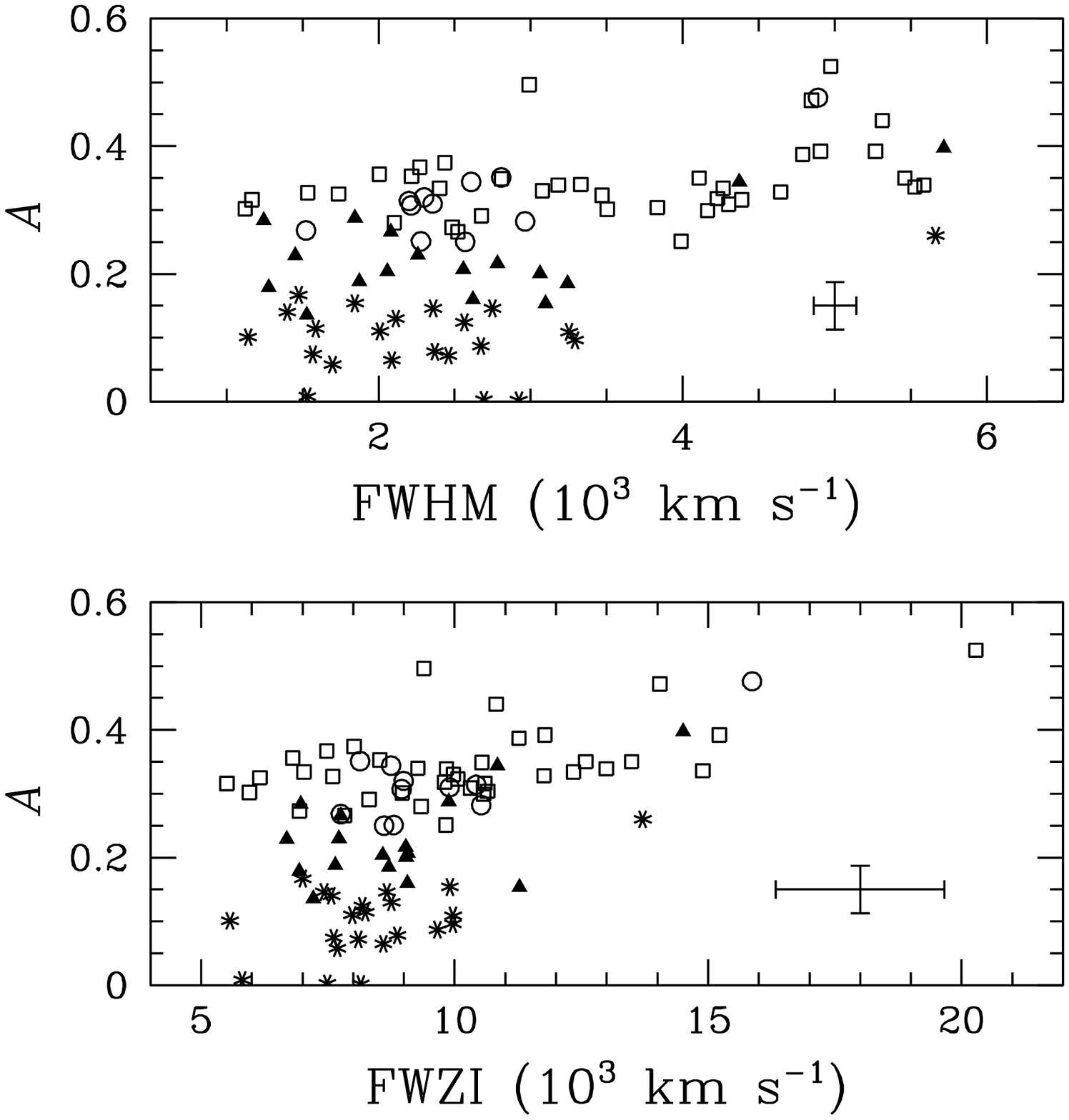}
\end{center}
\caption{{\bf Left:} Four examples of BP applied to the broad Balmer emission line components. {\bf Right:} The temperature parameter $A$ plotted vs. the FWHM (upper panel) and FWZI (lower panel) of the broad emission lines for the different BP results: asterisks for no BP fit; filled triangles for poor linear fit; open squares for good fits up to H$\delta$; open circles for fits up to H$\epsilon$. From La Mura et al. (2007). \label{f01}}
\end{figure}
If we take into account a series of transitions originated by a particular ion species, the normalized intensity of each line is a function of the upper level energy, that we can express as:
$$\log I_n = -\frac{\log e}{k_B T} E_{u 0} + const., \eqno(4)$$
where it has been assumed that the line emitting region is the same for the whole transition series. Eq.~(4) defines the Boltzmann Plot (BP) of the transition series. This formalism was originally developed for high density plasmas, but in La Mura et al. (2007) it was tested on a statistically relevant sample of type 1 AGNs, extracted from the 3$^{rd}$ data release of the {\it Sloan Digital Sky Survey} (SDSS). As it is illustrated in Fig.~\ref{f01}, the technique led to the identification different physical scenarios in the line emitting plasma. 

In general, it is observed that a satisfactory match with the underlying assumptions is achieved in approximately 30\%\ of the selected objects, with an improved statistics in the domain of sources with very broad emission lines. This result is also illustrated in Fig.~\ref{f01}, where the coefficient $A = -\log e / k_B T$, corresponding to the slope of the linear function defined in Eq.~(4), is compared with the width of the broad emission line components. In the range of narrow line emitting objects, instead, the BP finds indications of a higher degree of plasma ionization, probably due to a stronger interaction with the ionizing radiation field.

\sectionb{3}{OPTICAL LINE PROFILES}

The profiles of broad lines are influenced by the effects of complex kinematics within the source and of radiation transfer from the source to the observer. For this reason it is hardly conceivable that a simple analytic expression might be used to fit the outcome. If the line emitting region results from the combination of several kinematical components, multiple Gaussian functions provide reasonable fits to the observed profiles. However, the presence of distinct kinematical components may affect the profiles in a non-Gaussian form. A good way to estimate the importance of non-Gaussian contributions is to parameterize the observed line shapes by means of a Gauss--Hermite orthonormal expansion (see Van Der Marel \& Franx 1993). If we call $\alpha(v)$ the normal Gaussian function, the emission line profile in velocity units can be expressed as:
$$F(v) = F_0 \alpha(v - V_{sys}) \left[ 1 + \sum_{i = 3}^N h_i H_i(v - V_{sys}) \right], \eqno(5)$$
in which we call $F_0$ the profile normalization factor, $V_{sys}$ the systemic radial velocity offset between the line and the adopted reference frame, $H_i(v -− V_{sys})$ the $i^{\rm th}$ order Hermite polynomial, and $h_i$ the corresponding coefficient.

\begin{figure}[t]
\begin{center}
\includegraphics[width=6cm]{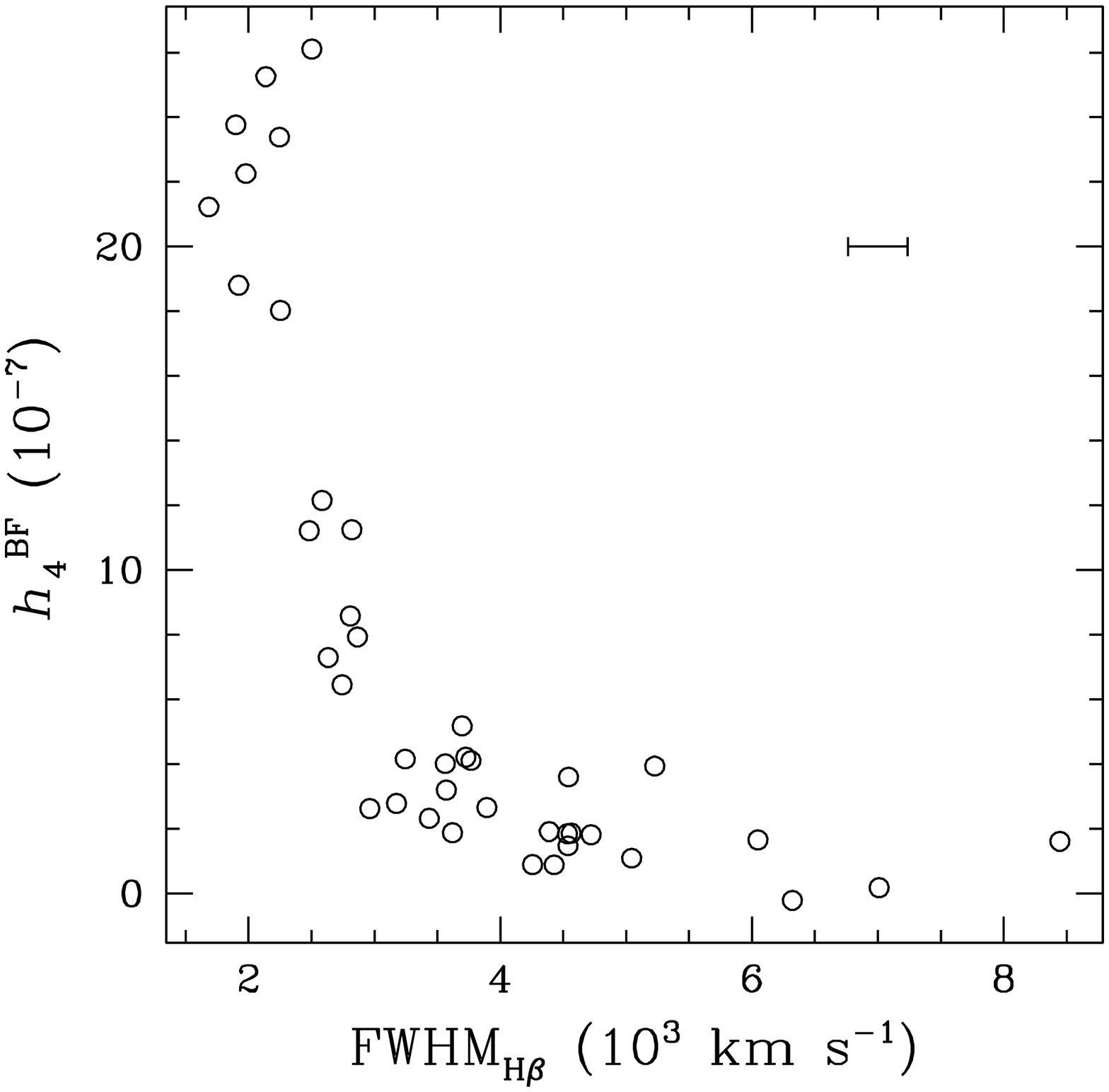}
\includegraphics[width=6cm]{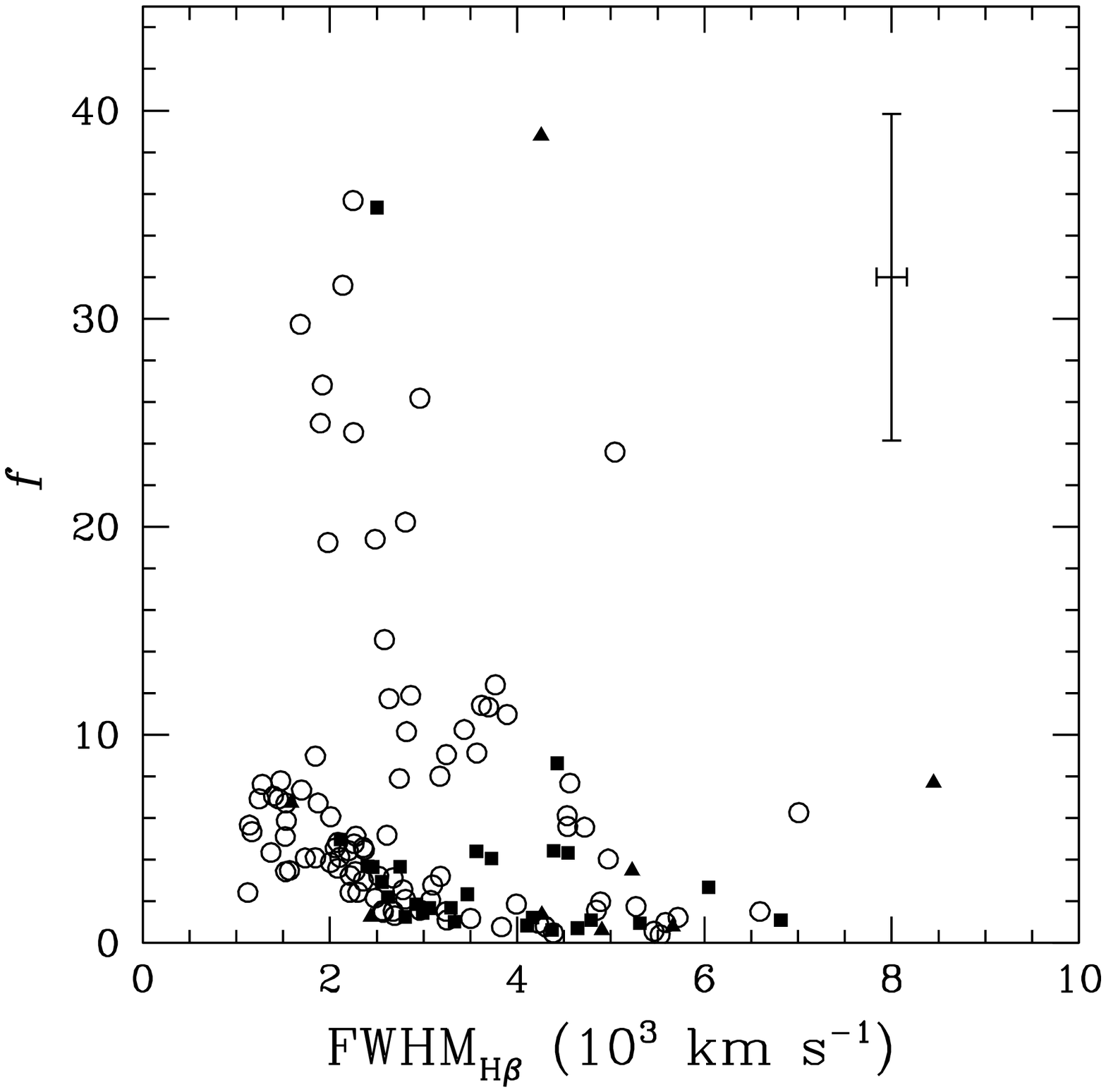}
\end{center}
\caption{{\bf Left:} Symmetric non-Gaussian profile components as function of FWHM$_{\rm H\beta}$; {\bf Right:} Geometrical factors accounting for the inclination of the rotating disk component plotted against FWHM$_{\rm H\beta}$. The strongest geometrical effects of flattening and inclination are observed in objects with FWHM$_{\rm H\beta} \leq 4000\, {\rm km\, s}^{-1}$. From La Mura et al. (2009). \label{f02}}
\end{figure}
As we show in Fig.~2, the symmetric component of the non-Gaussian contribution $h_4$  shows a remarkable decrease for increasing line profile width. Such an effect can be explained assuming that the BLR velocity field includes an ordered kinematical component, consistent with a flattened rotating structure, whose inclination implies a considerable variation of the geometrical factor $f$. Taking into account the role of the BLR inclination, in the formation of the observed line profiles, it is observed that the geometrical factor has an important influence on the determination of the BLR velocity field, but the effect is not critically relevant for narrow line emitting sources (FWHM$_{\rm H\beta} \leq 2000\, {\rm km\, s}^{-1}$), while it is important in broad line emitters ($2000\, {\rm km\, s}^{-1} \leq {\rm FWHM_{H\beta}} \leq 4000\, {\rm km\, s}^{-1}$).

\sectionb{4}{X-RAY SPECTRA}

The X-ray spectra considered in this work were collected with the different detectors of the EPIC camera, leading to the extraction of approximately simultaneous observations of the targets, with instruments having different energy responses. In this way we were able to exploit the technique of simultaneous spectral modeling, where it is required that a single model fits distinct data sets. To build the models, we used the XSPEC software (version 12.6.0q), fitting the spectra in several subsequent steps. At first, we ignored the low energy band data and we reproduced the spectrum between 2~keV and 10~keV by means of a simple power law, combined to a Gaussian function at a reference energy of 6.4~keV, accounting for the Fe~K$\alpha$ emission line. As a second step, we froze the power law component and we included in the fit the low energy band between 0.3~keV and 2~keV. This is a particularly complex region of the spectrum, where the contribution of unresolved emission and absorption features from highly ionized species can be relevant (Boller et al. 2007). Since the instrument performances did not provide the required significance to distinguish among the possible theoretical models that have been developed to account for the spectral energy distribution in this band, we added a combination of thermal components, using up to two contributions, depending on the resulting improvement of the fit. Once every component had a steady role in subsequent iterations, we allowed for the fit of the whole spectrum. The final model had to account for the X-ray absorption expected from the neutral gas column density along the observation direction within our Galaxy.

\begin{table}[t]
\caption{Measured X-ray and optical spectral properties. \label{t01}}
\begin{center}
\begin{footnotesize}
\begin{tabular}{lcccc}
\hline
\hline
Name & FWHM$_{{\rm H}\beta}$ & $E_{Fe}$ & $EW_{Fe}$ & $\Gamma$\\
 & (km s$^{-1}$) & (keV) & (eV) & \\
\hline
2MASX J03063958+0003426 & 1949 $\pm$ 170 & 6.59 $\pm$ 0.18 & 202.43 $\pm$ 88.31 & 1.87 $\pm$ 0.02 \\
MCG+04-22-042 & 1946 $\pm$ 170 & 6.52 $\pm$ 0.06 & 95.55 $\pm$ 48.14 & 1.74 $\pm$ 0.04 \\
Mrk 110 & 2339 $\pm$ 348 & 6.45 $\pm$ 0.04 & 59.57 $\pm$ 16.68 & 1.71 $\pm$ 0.01 \\
PG 1114+445 & 5090 $\pm$ 462 & 6.42 $\pm$ 0.05 & 115.47 $\pm$ 31.05 & 1.53 $\pm$ 0.05 \\
PG 1115+407 & 1760 $\pm$ 170 & 7.18 $\pm$ 0.10 & 212.75 $\pm$ 130.91 & 2.23 $\pm$ 0.06 \\
2E1216+0700 & 2028 $\pm$ 170 & 6.48 $\pm$ 0.42 & 412.80 $\pm$ 276.98 & 2.21 $\pm$ 0.05 \\
Mrk 50 & 5540 $\pm$ 199 & 6.37 $\pm$ 0.06 & 61.53 $\pm$ 40.96 & 1.79 $\pm$ 0.03 \\
Was 61 & 1591 $\pm$ 170 & 6.56 $\pm$ 0.11 & 95.86 $\pm$ 42.68 & 2.03 $\pm$ 0.03 \\
PG 1352+183 & 4458 $\pm$ 829 & 6.45 $\pm$ 0.36 & 179.48 $\pm$ 116.27 & 1.92 $\pm$ 0.06 \\
Mrk 464 & 6188 $\pm$ 575 & 6.43 $\pm$ 0.18 & 192.70 $\pm$ 117.88 & 1.54 $\pm$ 0.06 \\
PG 1415+451 & 2850 $\pm$ 170 & 6.67 $\pm$ 0.07 & 163.30 $\pm$ 92.51 & 1.97 $\pm$ 0.05 \\
NGC 5683 & 5273 $\pm$ 313 & 6.41 $\pm$ 0.20 & 514.44 $\pm$ 336.87 & 1.74 $\pm$ 0.15 \\
Mrk 290 & 4202 $\pm$ 170 & 6.37 $\pm$ 0.06 & 74.51 $\pm$ 35.30 & 1.54 $\pm$ 0.02 \\
Mrk 493 & 988 $\pm$ 170 & 6.86 $\pm$ 0.19 & 239.25 $\pm$ 155.68 & 2.10 $\pm$ 0.05 \\
\hline
\end{tabular}
\end{footnotesize}
\end{center}
\end{table}

\sectionb{5}{CONCLUSIONS}

\begin{figure}[t]
\begin{center}
\includegraphics[width=6cm]{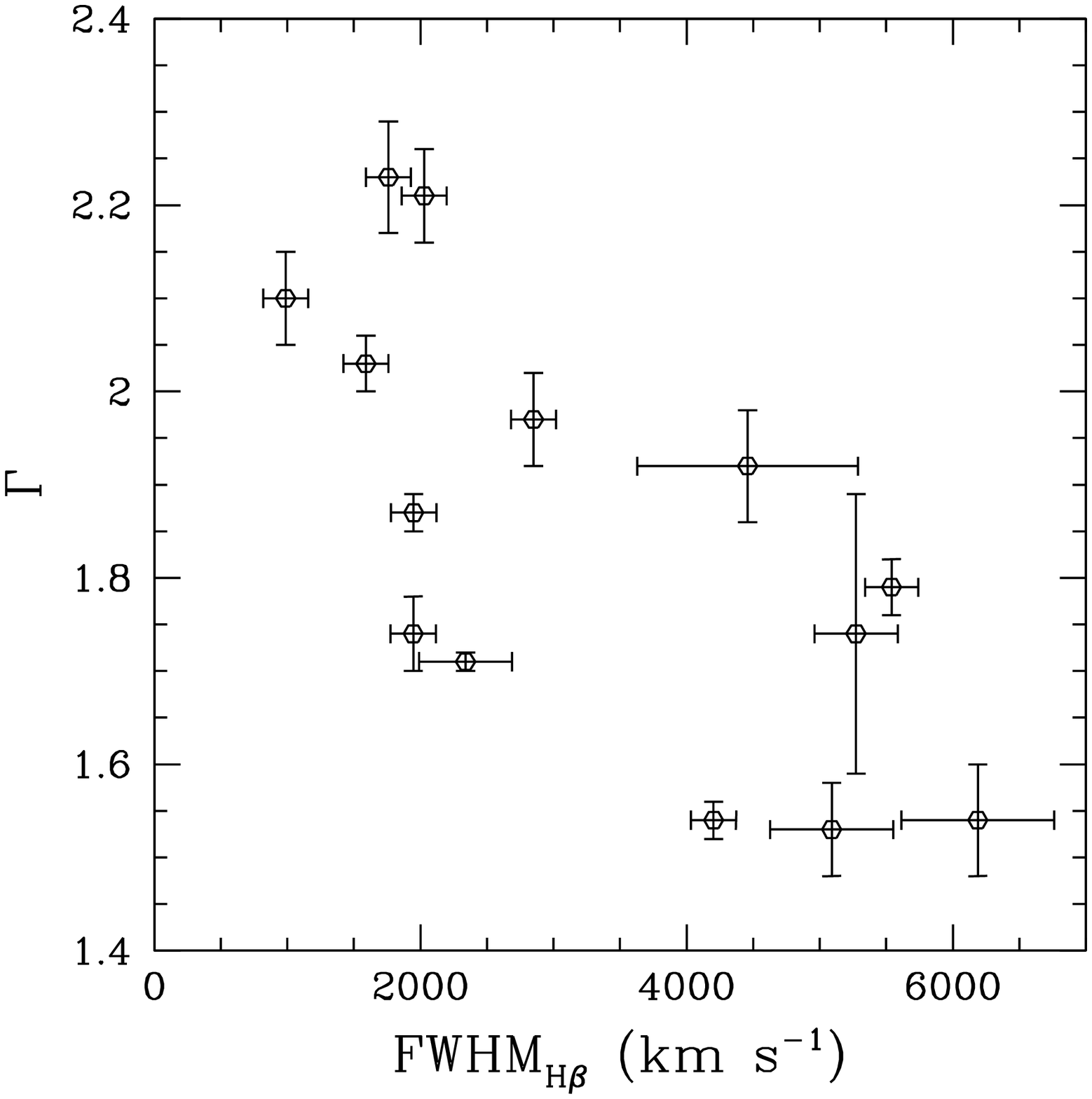}
\includegraphics[width=6cm]{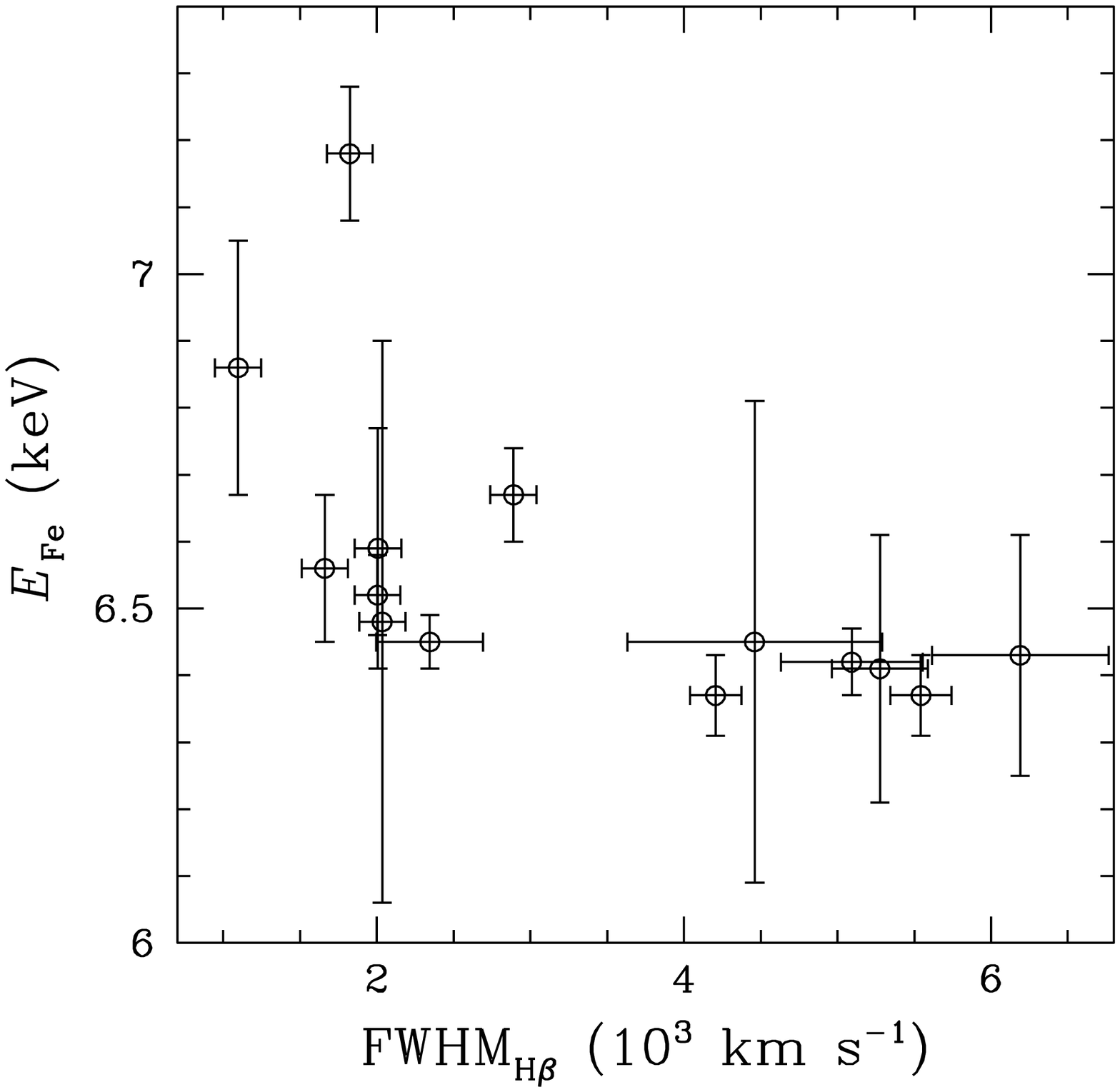}
\caption{Hard X-ray power law continuum photon index $\Gamma$ (left panel) and the Fe line central energy (right panel) plotted as functions of FWHM$_{{\rm H}\beta}$. \label{f03}}
\end{center}
\end{figure}

The most notable results of the comparison among optical and X-ray spectral properties are illustrated in Fig.~\ref{f03}. As it was already noted (e. g. Shemmer et al. 2008), there is a clear increase of the hard X-ray power law slope moving from broad to narrow line emitters. In addition, Sulentic et al. (2008) noted similar differences in the slope of the power law X-ray continuum, which fit in the distinction among population A and B AGNs. On the other hand, reflection of the hard X-ray continuum very close to the central source originates a blend of Fe lines, with a resulting energy that depends on the ionization degree of the medium (Ross et al. 1999). The most prominent contribution to this component descends from the Fe~$K\alpha$ emission of the variously ionized Fe atoms. The observed energy of the blend shows a steep raise in the domain of narrow line emitting sources. Indeed, an indication of such an effect in the case of Narrow Line Seyfert 1 galaxies (NLS1s) has been previously given in Romano et al. (2002), who suggested a reflection within a significantly ionized medium as a possible interpretation.

In this work, we studied the Fe line central energies to investigate the properties of the reflecting medium within a sample of targets including both narrow and broad line emitters. The results that we achieved so far suggest that, while in the range of very broad line emitting sources we observe radiation coming from a neutral, unperturbed component of the medium, with little evidence for a strong orientation dependence, the case of narrow line emitting sources is different. These objects, indeed, are characterised by a higher energy reflection line, suggesting either the contribution of an ionized medium or the presence of powerful outflows, such as the ones that could be expected in the case of strong winds. Although we are not able to derive any unambiguous conclusion from the properties of the observed objects, the analysis of X-ray spectra confirms the existence of a perturbed medium in the central regions of narrow line emitting AGNs, with respect to the situation more commonly observed in broad line sources. This result fits appreciably well with the analysis of the BLR physical conditions, carried out in the optical domain, supporting the idea that narrow emission line sources are more likely to be powered by relatively low mass black holes, accreting at very high rates.

\thanks{We gratefully aknowledge the LOC and SOC for the invitation and for the opportunity to present this contribution.}

\References

\refb Boller, T., Balestra, I., Kollatschny, W. 2007, A\&A, 465, 87
%\refb Chen, H., Wang, J.-M., Ho, L. C., Chen, Y.-M, Bian, W.-H., Xue, S. J. 2008, ApJ, 683, L115
\refb La Mura, G., Popovi\'c, L. \v{C}., Ciroi, S., Rafanelli, P., Ili\'c, D. 2007, ApJ, 671, 104
\refb La Mura, G., Di Mille, F., Ciroi, S., Popovi\'c, L. \v{C}., Rafanelli, P. 2009, ApJ, 693, 1437
\refb Popovi\'c, L. \v{C}. 2003, ApJ, 599, 140
%\refb Popovi\'c, L. \v{C}., Mediavilla, E., Bon, E., Ili\'c, D. 2004, A\&A, 423, 909
\refb Popovi\'c, L. \v{C}. 2006, ApJ, 650, 1217 (an Erratum)
\refb Romano, P., Turner, T. J., Mathur, S., George, I. M. 2002, ApJ, 564, 162
\refb Ross, R. R., Fabian, A. C., Young, A. J. 1999, MNRAS, 306, 461
\refb Shemmer, O., Brandt, W. N., Netzer, H., Maiolino, R., Kaspi, S. 2008, ApJ, 682, 81
\refb Sulentic, J. W., Zamfir, S., Marziani, P., Dultzin, D. 2008, RevMexAA, 32, 51
\refb Van Der Marel, R. \& Franx, M. 1993, ApJ, 407, 525

\end{document}